\documentclass[12pt]{article}
\usepackage{latexsym}
\usepackage{amssymb}
\usepackage{amsmath}
\usepackage{graphicx}

\begin{document}

\title{Thermodynamics and Evaporation of Closed Black Cosmic Strings}
\author{\small Andrew DeBenedictis \footnote{e-mail:adebened@sfu.ca}
\\ \it{\small Department of Physics}
\\ \it{\small Simon Fraser University, Burnaby, B.C., Canada V5A 1S6}}
\date{July 2, 1998}
\maketitle

\begin{abstract}
Both the canonical and microcanonical ensembles are utilized to study the 
thermodynamic and evaporation properties of a closed black cosmic string  
whose spacetime is asymptotically anti deSitter. There are similarities 
and differences to the Schwarzschild-anti deSitter and 2+1 BTZ black hole 
solutions. It is found that there exist regimes of black string/thermal 
radiation equilibrium as well as stable remnant regimes. The relevance 
to black hole evaporation is discussed.
\end{abstract}

\section{Introduction}
\qquad The interesting field of black hole thermodynamics began with the 
discovery by Hawking ~\cite{hawkevap} that black holes actually radiate 
energy as if they were hot bodies. The temperature being governed by a 
multiple of the surface gravity, $\kappa$, of the black hole. 
It is now also 
well know that the area of the event horizon is a direct measure of the 
entropy of the black hole ~\cite{bekentrop} by the relation $S=G^{-1}A/4$, 
$A$ being the area of the event horizon.

\qquad The above properties were originally derived in spacetimes which are 
asymptotically space-like flat. It is useful, however, to study such 
phenomena in spacetimes which are not necessarily asymptotically flat 
since it is unknown how good the assumption of asymptotic flatness may 
be. Black hole solutions also exist which tend either to deSitter 
spacetime (if the cosmological constant is positive) or anti-deSitter 
(adS) (if the cosmological constant is negative). 

\qquad These studies have been extended to the Schwarzschild-deSitter 
 ~\cite{schwds} as well as the Schwarzschild-adS case 
~\cite{handpads} (the latter being of particuilar relevance to the study 
here). It was found that an identical area-entropy law holds in the 
asymptotically adS case as does in the case of asymptotic flatness.

\qquad Another solution which tends to adS is the 2+1 dimensional black 
hole formulated by Banados, Teitelboim and Zanelli ~\cite{btzorig} 
(hereafter referred to as the BTZ black hole). This must be the case in 
lower dimensional gravity since for $D<4$, $D$ being the number of 
spacetime dimensions, the Riemann curvature tensor must vanish if the 
Ricci tensor is zero. Thermodynamic studies of 
this system have also been done \cite{btzorig}, \cite{reznik}, \cite{oda}, 
~\cite{mannthermo}. The entropy in this case turns out to be measured by 
the circumference of the event horizon.

\qquad The metric of the spacetime will be flat torus model developed by 
Lemos and Zanchin ~\cite{lemos1} with $S^1\times S^1$ topology. Black 
cosmic strings have also been studied by Kaloper 
~\cite{kaloper}. The metric is given by \footnote{units are used here such 
that $c=k_B=\hbar=1$. This gives the Planck mass a value of 
$m_p=G^{-\frac{1}{2}}$.}   
(without charge or angular momentum) 
\begin{equation}
ds^2=-\left(\alpha^{2}\rho^{2}-\frac{2GM}{\pi\rho}\right)\,dt^2+
\frac{d\rho^2}{\left(\alpha^2\rho^2-\frac{2GM}{\pi\rho}\right)}\,+
\rho^2\left(d\varphi^2+d\vartheta^2\right), \label{eq:metric}
\end{equation}
where $M$ is the total mass of the string and 
$\alpha=-\frac{1}{3}\Lambda$ ($\Lambda$ being the cosmological constant).
The coordinate ranges are as follows: $-\infty < t < +\infty$, $0\leq \rho < 
+\infty$, $0\leq \varphi < 2\pi$ and $0\leq \vartheta < 2\pi$. By the 
redefinition $\vartheta\rightarrow \alpha z$ with $-\infty \leq z < 
+\infty$ the above metric describes an infinitely long black cosmic 
string with a mass per unit length of $\alpha M/2\pi$. The Kretschmann 
scalar is given by 
\begin{equation} R_{\alpha \beta \gamma \delta}R^{\alpha \beta \gamma 
\delta}=24\alpha^2\left(1+ \frac{(2GM)^2}{2\pi^2\alpha^4\rho^6}\right)
\end{equation}
from which it can be seen that a true polynomial singularity exists at 
$\rho=0$. An event horizon is located at $\rho_{H}= 
\left(\frac{2GM}{\pi\alpha^2}\right)^{\frac{1}{3}}$.

\qquad Since the background reference spacetime is adS, special boundary 
conditions must be imposed at the time-like surface $\rho=\infty$ so that 
a well defined Cauchy problem will exist ~\cite{isham} (see Fig.1). The 
standard ``reflective" boundary condition will be used here for massless 
particles noting that massive particles do not reach spatial infinity. As 
pointed out in ~\cite{reznik}, this has some interesting effects on black 
hole evaporation as will be discussed below.

\qquad Both the canonical and $\mu$icrocanonical ensembles will be used in
this study. Since reflection is imposed at infinity and the metric is 
static, energy is conserved. The total amount of energy is also bounded
due to the presence of the cosmological constant which causes large
(infinite) redshift effects away from the origin ~\cite{allen}.

\begin{figure}[ht]
\vspace{0.5in}
\label{fig1}
\includegraphics[bb=71 339 305 738, width=0.3\textwidth,clip]{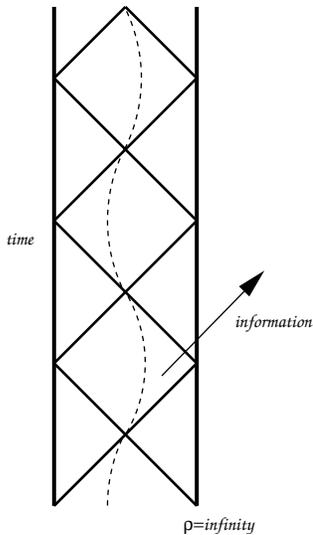}
\caption{\small Penrose diagram for adS spacetime. The time-like 
surface allows null information to ``leak out" at spatial infinity (arrow) 
therefore a reflective condition must be enforced at this surface. The 
dotted line displays a representative time-like geodesic which is 
automatically reflected back in the spacetime.} 
\vspace{0.5in}
\end{figure}   

\section{Canonical Ensemble}
\qquad Since there is no time periodicity in its Euclidean extension, adS 
spacetime has no natural temperature. Thermal states can be defined, 
however, by imposing a periodicity of $\beta=T^{-1}$ in imaginary time 
where $T$ is the desired temperature. The resulting states will have a 
local temperature $T_{local}$;
\begin{equation}
T_{local}=\frac{\beta^{-1}}{g_{00}^{1/2}}.
\end{equation}

Unlike the asymptotically flat 
case, the canonical ensemble can be perfectly  defined in a spacetime 
which is adS \footnote{for a review of the problems associated 
with defining a canonical ensemble in asymptotically flat spacetimes see 
~\cite{handpads}}. The stress-energy tensor for a conformally coupled 
scalar field is given by ~\cite{handpads}~\cite{allen}
\begin{equation}
T^{\mu}_{\nu}=\frac{\pi^2}{90} 
\frac{gT^{4}}{\left(1+\alpha^2 r^2\right)^{2}}\left(\delta^{\mu}_{\nu}- 
4\delta^{\mu}_{0}\delta^{0}_{\nu}\right) + O(\alpha^{-2}T^2),
\end{equation}
where $g$ is the number of spin states.
A mass integral can be formed by contraction with a properly normalised 
time-like killing vector. Integrating over all space gives the total 
energy of the thermal state,
\begin{equation}
E\approx\frac{\pi^4}{30}gT^{4}\alpha^{-3}.
\end{equation}
The integral of this quantity with respect to inverse temperature, 
$\beta$, yields the partition function $Z$.
\begin{equation}
ln(Z)=\frac{\pi^4}{90}\frac{g}{(\alpha\beta)^3} + O(\alpha^{-1}\beta^{-1}),
\end{equation}
from which the free energy may be calculated as
\begin{equation}
F=-T\,ln(Z)=-\frac{\pi^4}{90}\frac{g}{\alpha^3\beta^4} + 
O(\alpha^{-1}\beta^{-2}). \label{eq:freerad} 
\end{equation}
It should be noted that gravitational back reaction effects of the 
thermal radiation have been ignored. There is also a temperature above 
which the radiation will be unstable to collapse forming a black hole. 
This temperature is given by ~\cite{handpads}
\begin{equation}
T_{collapse}\approx (gG)^{-\frac{1}{4}}\alpha^{\frac{1}{2}}, 
\label{eq:collapset} 
\end{equation}
and can be derived from the following argument. If the radiation is 
spherically symmetric, the Einstein field equations give
\begin{equation}
g^{11}=A+\alpha^2 r^2-\frac{8\pi G}{r}\int^{r}{\varepsilon r^2\,dr}
\end{equation}
where $A$ is a constant and $\varepsilon$ is given by 
\begin{equation}
\varepsilon=\frac{\pi^2}{30}gT^4.
\end{equation}
A horizon will form unless condition (\ref{eq:collapset}) holds.

\qquad Attention is now turned to the black string spacetime. The 
Euclidean extension is obtained by making the transformation $\tau=\imath 
t$ in (\ref{eq:metric}) and the Hawking temperature can be 
calculated by demanding that the Euclidean metric be regular on the 
horizon. This gives a periodicity in imaginary time
\begin{equation}
\beta=T^{-1}=\frac{2}{3(GM)^{\frac{1}{3}}} 
\left(\frac{2\pi^2}{\alpha^2}\right)^{\frac{2}{3}}.
\end{equation} 
The expectation value of the energy is given by the total mass, $M$, of 
the black string \footnote{Since the straight cosmic string is infinite, 
some quantities such as total energy are unbounded. This leads to infinite 
quantities when calculating certain thermodynamic functions. 
Intensive quantities, however, are still well defined.} so that the heat 
capacity $\partial M/\partial T$ is always positive and black hole states 
may be in thermal equilibrium with radiation. Note that unlike the 
Schwarzschild case (whose temperature is inversly proportional to the 
black hole mass) the temperature here is proportional to $M^{1/3}$. For 
the BTZ black hole temperature is proportional to $M^{1/2}$.

\qquad From the above properties one easily obtains the partition 
function 
\begin{equation}
ln\left(Z\right)=\frac{16}{27}\frac{\pi^4}{G\alpha^4}\beta^{-2} 
\label{eq:lnz} \end{equation}
and the free energy
\begin{equation}
F=-\frac{16}{27}\frac{\pi^4}{G\alpha^4}\beta^{-3}=-\frac{M}{2}. 
\label{eq:freebstr} 
\end{equation}
By comparing free energies (\ref{eq:freerad}) and (\ref{eq:freebstr}) 
it is noted that for $T > 160/(3\alpha gG)$ the radiation free energy is 
less than the black string free energy. A black string configuration with 
a temperature higher than this will therefore totally evaporate to a 
radiation state. This crossover occurs at a mass of
\begin{equation}
M\approx\frac{1.31\times 10^8}{729}\frac{\pi^4}{\alpha^{7} gG^4}.
\end{equation}
However, since above temperature (\ref{eq:collapset}) the radiation 
self gravitation will cause it to collapse, only black 
hole solutions are stable if
\begin{equation}
\alpha < \left(\frac{160}{3}\right)^{2/3}(gG)^{-1/2}.
\end{equation}

\qquad The entropy of the black string is given by
\begin{eqnarray}
S&=&\frac{M}{T}-\frac{F}{T} \nonumber \\
&=&\frac{16}{9}\frac{\pi^4}{G\alpha^4}T^2 \nonumber \\
&=& \frac{\pi}{G}A \label{eq:entropy}
\end{eqnarray}
where $A$ is the cross sectional area of a disk of radius 
$\rho=\rho_{H}$. (\ref{eq:entropy}) is therefore the analog of the 
entropy/area relation for spherically symmetric systems. This result 
shows that the canonical ensemble is well defined in the black string 
spacetime. The defining integral for the partition function is
\begin{equation}
Z=\int{N(M)e^{-M\beta}\,dM}, \label{eq:defpart}
\end{equation}
where $N(M)$ is the density of states given by
\begin{equation}
N(M)\propto e^{S}=e^{\propto M^{2/3}}
\end{equation}
so that (\ref{eq:defpart}) is well defined. In asymptotically flat 
spacetimes the density of states goes as ~\cite{handpads} $exp(\propto 
M^2)$ so that (\ref{eq:defpart}) does not converge.

\section{Microcanonical Ensemble}
\qquad adS, unlike asymptotically flat, spacetime does not 
require one to introduce an artificial ``box" to bound the system. This is 
due to the fact that geodesics in adS automatically reflect massive 
particles. Particles of zero mass, which do propagate to spatial 
infinity, will be reflected by the appropriate boundary conditions 
(see Fig.1).

\qquad The density of states is given by the inverse LaPlace transform 
\begin{equation}
N(E)=-\frac{\imath}{2\pi} 
\int_{-\imath\infty}^{+\imath\infty}{Z(\beta)e^{\beta E}d\beta}. 
\label{eq:microdense}
\end{equation}
For adS the quantities of interest have been computed in ~\cite{handpads} 
and are quoted here. For stable thermal radiation
\begin{equation}
Z\approx exp\left(\frac{\pi^4}{30}g(\alpha\beta)^{-3}\right).
\end{equation}
A saddle point to (\ref{eq:microdense}) exists at
\begin{equation}
\beta\approx\left(\frac{\pi^4}{30}g\alpha^{-3}E^{-1}\right)^{1/4}
\end{equation}
so that in the stationary phase approximation the number of states is 
given by
\begin{equation}
N(E)\approx exp\left[\frac{4\pi}{3} 
\left(\frac{g}{30\alpha^3}\right)^{1/4}E^{3/4}\right].
\end{equation}
\qquad For the black string spacetime recall that $Z(\beta)$ is given by 
(\ref{eq:lnz}) so that a saddle point exists at
\begin{equation}
\beta=\left(\frac{32}{27}\frac{\pi^4}{G\alpha^4}\right)^{1/3}E^{-1/3}
\end{equation}
yielding
\begin{equation}
N_{black string}(E)=exp\left[\frac{3}{2} 
\left(\frac{32\pi^4}{27G\alpha^4}\right)^{1/3}E^{2/3}\right].
\end{equation}
The energy dependence $e^{\propto E^{2/3}}$ is similar to the 
Schwarzschild-adS black hole~\cite{handpads} for large mass. The BTZ black 
hole has an
$e^{\propto E^{1/2}}$ dependence~\cite{reznik}. From this relation it is 
determined that $N_{radiation}>N_{black string}$ when 
\begin{equation}
E>\sim\frac{1.8\times 10^6}{8.2}\frac{\pi^4}{G^{4}\alpha^{7}g^{3}}.
\end{equation}

\qquad At the stationary phase point the temperature/energy relation of the 
system is given by
\begin{equation}
E=M_{black string}+E_{radiation}\approx 
\frac{32\pi^4}{27G\alpha^4}\beta^{-3}+ 
\frac{\pi^4}{30}g\alpha^{-3}\beta^{-4}. \label{eq:etot}
\end{equation}
Therefore, when the radiation contribution is significant and there 
exists a state in which the black string is in thermal equilibrium with 
thermal radiation, the number of states is given by
\begin{equation}
N(E)\approx exp\left[\left(\frac{\pi^2}{\alpha^2} 2 
M\right)^{2/3}G^{-1/3}+ 
\frac{4\pi}{3}\left(\frac{g}{\alpha^3 30}\right)^{1/4}\,E_{radiation}^{3/4} 
\right].
\end{equation}

\section{Evaporation of the Black String}
\qquad It is interesting to speculate about the effects of evaporation in 
a spacetime which is asymptotically adS. An excellent discussion of the 
effects for the BTZ black hole can be found in ~\cite{reznik}. The 
central question is whether or not mass loss from the black string can 
occur since, unlike the asymptotically flat case, massive particles 
automatically return to their original position on a time scale governed 
by the cosmological constant. Massless particles are reflected back at 
spatial infinity by the boundary condition on a similar timescale. 
Complete evaporation is therefore not possible in all situations.

\qquad For situations where total evaporation is possible there are several 
possibilities. There is the situation where evaporation is complete and 
the remnant will be pure radiation with no black string. The other 
possibility is the case where 
$\alpha<\left(\frac{160}{3}\right)^{2/3}(gG)^{-1/2}$. In this case the 
resulting 
radiation is unstable to collapse and will therefore form a black hole. 
Since it is unlikely that the recollapsing matter will form a toroidal 
black hole the initial configuration again will have completely 
evaporated, the remnant most likely being a spherical black hole. It is 
interesting to speculate on the significance of such evaporation in the 
context of cosmic string theory. 

\qquad It is well know from the theory of cosmic strings in flat 
spacetime ~\cite{vilenkin} that oscillating string loops will lose 
energy\footnote{It is assumed that the cosmic 
string is not a superconducting or global string in which case there may 
be other significant methods of energy loss.} via gravitational wave 
emission. This occurs at a rate approximately given by
\begin{equation}
\dot{E}=G\,K\,\mu^2
\end{equation}
where $K$ is a geometric factor and $\mu$ is the string tension. The 
lifetime of a loop losing energy via this mechanism is approximately
\begin{equation}
t\approx\frac{L}{KG\mu}
\end{equation} 
$L$ being the length of the loop. The static toroidal black string 
studied here admits another mechanism for decay.

\section{Acknowledgments}
The author would like to thank Dr. K. S. Viswanathan for helpful advice on 
this and related projects. Discussions on Relativity with Dr. A. Das are 
always enlightening. This work was partially funded by an NSERC operating 
grant.

\newpage
\bibliographystyle{unsrt}

\begin{thebibliography}{10}

\bibitem{hawkevap}
Hawking~S~W
\newblock 1975 {\em Commun. Math. Phys.} {\bf 43} 199

\bibitem{bekentrop}
Bekenstein~J~D
\newblock 1973 {\em Phys. Rev. D} {\bf 7} 2333

\bibitem{schwds}
Gibbons~G~W and Hawking~S~W
\newblock 1977 {\em Phys. Rev. D} {\bf 15} 2738

\bibitem{handpads}
Hawking~S~W and Page~D~N
\newblock 1983 {\em Commun. Math. Phys.} {\bf 87} 577

\bibitem{btzorig}
Banados~M, Teitelboim~C and Zanelli~J
\newblock (1992) {\em Phys. Rev. Lett.} {\bf 69} 1849

\bibitem{reznik}
Reznik~B
\newblock 1995 {\em Phys. Rev. D} {\bf 51} 1728

\bibitem{oda}
Oda~I
\newblock 1997 {\em Phys. Lett.} {\bf 409} 88

\bibitem{mannthermo}
Brown~J~D, Creighton~J and Mann~R~B
\newblock 1994 {\em Phys. Rev. D} {\bf 50} 6394

\bibitem{lemos1}
Lemos~J~P~S and Zanchin~V~T
\newblock 1996 {\em Phys. Rev. D} {\bf 54} 3840

\bibitem{kaloper}
Kaloper~N
\newblock 1993 {\em Phys. Rev. D} {\bf 48} 4658

\bibitem{isham}
Avis~S~J, Isham~C~J and Storey~D
\newblock 1978 {\em Phys. Rev. D} {\bf 18} 3565

\bibitem{allen}
Allen~B, Folacci~A and Gibbons~G~W
\newblock 1987 {\em Phys. Lett. B} {\bf 189} 304

\bibitem{vilenkin}
Vilenkin~A and Shellard~E~P~S 
\newblock 1994 {\em Cosmic Strings and other Topological Defects} 
(Cambridge: Cambridge University Press)

\end{thebibliography}

\end{document}